\documentclass[a4paper,twocolumn,nofootinbib]{revtex4}
\usepackage{amsmath,epsfig}
\usepackage{amssymb}

\begin{document}

\title{Resummation of projectile-target
multiple scatterings and parton saturation}

\author{S. Munier}
\author{F. Schwennsen}
\affiliation{Centre de physique th{\'e}orique, \'Ecole polytechnique,
CNRS, 91128~Palaiseau, France}


\begin{abstract}
In the framework of a toy model which possesses the main
features of QCD in the high energy limit,
we conduct a numerical study of scattering amplitudes
constructed from parton splittings and projectile-target
multiple interactions, in a way that unitarizes the amplitudes 
without however explicit saturation in the wavefunction of the
incoming states.
This calculation is performed in
two different ways. One of these formulations, the closest to field theory,
involves the numerical resummation
of a factorially divergent series, for which we develop appropriate
numerical tools.
We accurately compare the properties of the resulting
amplitudes with what would
be expected if saturation were explicitly included in the
evolution of the states.
We observe that the amplitudes have similar properties 
in a small but finite range of rapidity in the beginning of 
the evolution, as expected.
Some of the features
of reaction-diffusion processes are already present
in that range, 
even when saturation is
left out of the model.
\end{abstract}

\maketitle

\section{Introduction}

Saturation of parton densities
is a phenomenon that is expected to occur in the 
scattering of very
fast hadrons \cite{GLR,MQ}.
It is the statement that the number of partons
per unit of transverse 
phase space in the Fock states of the incoming hadrons
does not grow exponentially forever with energy (or rapidity), 
as would come out of
a naive solution
of linear evolution equations: The growth gets softer
at very high energies,
in such a way that the unitarity of the probabilities
of scattering be preserved.

However, it seems very difficult to get saturation 
from a field-theoretical calculation.
So far, the problem has not even been formulated properly
in QCD, for it is already very challenging to identify the graphs
that should be taken into account.
Equations such as the Balitsky equation \cite{Balitsky:1995ub} (which is
in fact a hierarchy of equations)
have been written down, but 
they do not exhibit saturation in an obvious way (if at all),
and they are anyway extremely difficult to solve.

Despite these fundamental difficulties, 
new quantitative results could
be obtained for QCD amplitudes in the saturation 
regime over the last few years \cite{Mueller:2004sea,IMM}.
In short, they rely on an analogy with reaction-diffusion processes
\cite{IMM}
but not on the computation of definite Feynman graphs.
The obtained results do not depend on the way how partons saturate,
but they seem to 
require that there be such saturation phenomena.
So saturation was assumed rather than found in these approaches.
The matching of this statistical picture
with field-theory calculations
has been attempted \cite{Iancu:2004iy}, and has led to the statement that 
the Balitsky-Jalilian Marian-Iancu-McLerran-Weigert-Leonidov-Kovner 
(B-JIMWLK) equations \cite{Balitsky:1995ub,JIMWLK,Weigert:2005us}
were incomplete, and that
they had to be supplemented with new terms.
The problem was identified as follows: The Balitsky equations
only contain splittings of partons, while mergings are needed
in order to achieve saturation.
The hope that merging rates could be obtained by 
boosting splitting vertices was turned down by the finding
that this procedure would lead to negative 
probabilities \cite{MSW,IST}, 
which is at least inconvenient in practice \cite{MMX},
if not completely meaningless.

In this paper, we go back to the original formulation 
of saturation by Mueller
in the context of the color dipole model \cite{M1,M2,M3}, 
which was in fact
based on the assumption that saturation of the parton densities is
equivalent to unitarization of the scattering amplitudes
through multiple exchanges between linearly-evolved Fock
states of the incoming particles (if the rapidity is not
too high).
Analytical calculations have been achieved within similar
approximations (see Refs.~\cite{Levin1,Levin2} for recent progress),
but the result looks always complicated in QCD and thus
difficult to play with and to interpret.
Numerical studies were conducted by Salam \cite{S1,S_MC,MS},
but at that time, there was no good theoretical understanding
of the properties that scattering amplitudes
should exhibit at high energy.
In the light of our present knowledge of 
saturation, that
enables us to characterize saturation
by analyzing the properties of some traveling waves,
we evaluate numerically, on toy models, how good this procedure
is when only splittings and multiple exchanges are allowed.
Another important highlight of the present work is that we
are able to resum numerically the asymptotic series
that can be constructed
out of an expansion in the number of rescatterings, and which
has the structure of the proper field-theoretical
series of successive Pomeron exchanges.

\begin{figure}
\begin{center}
\epsfig{file=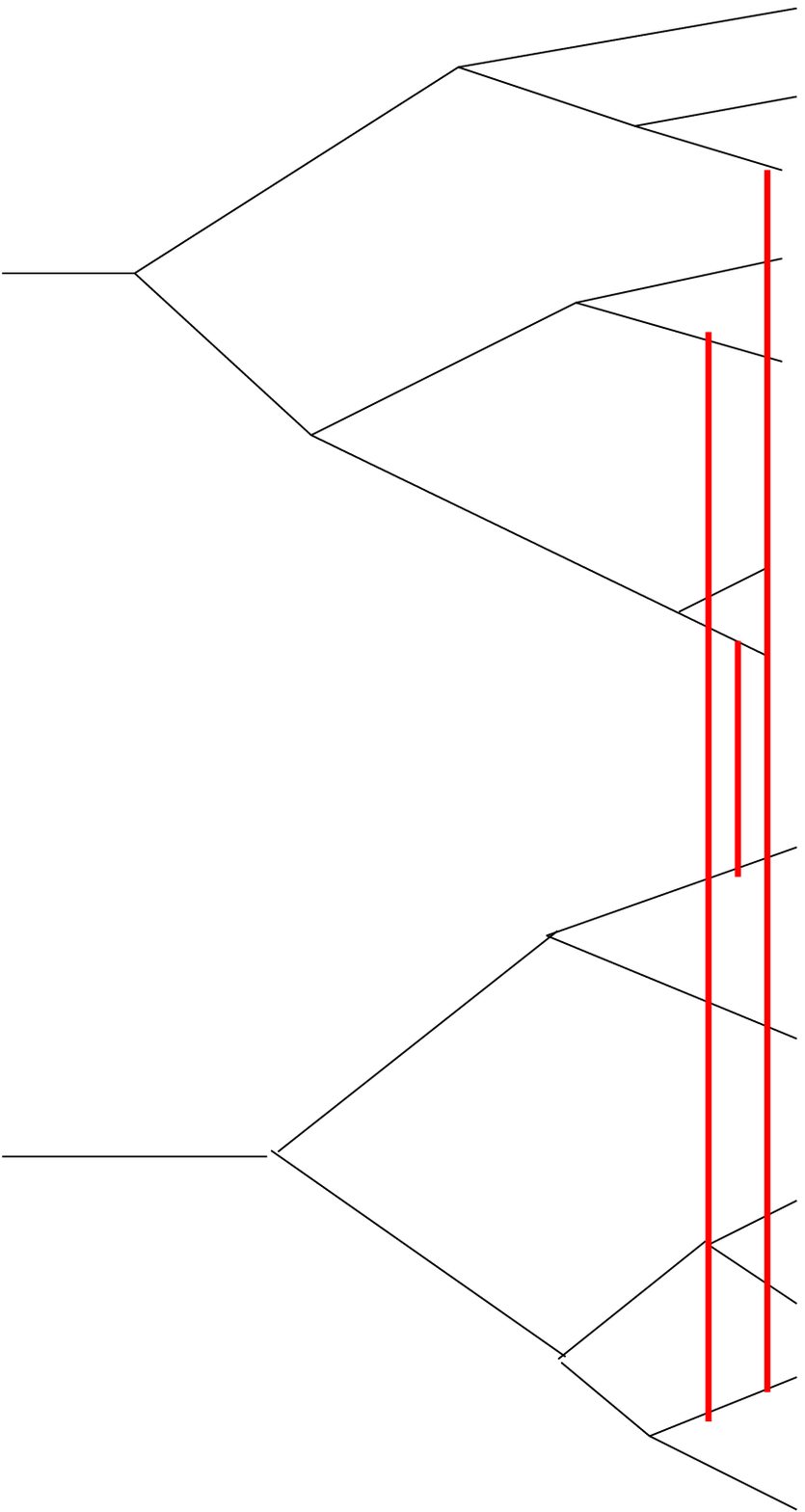,width=0.15\textwidth}\hskip 0.1\textwidth
\epsfig{file=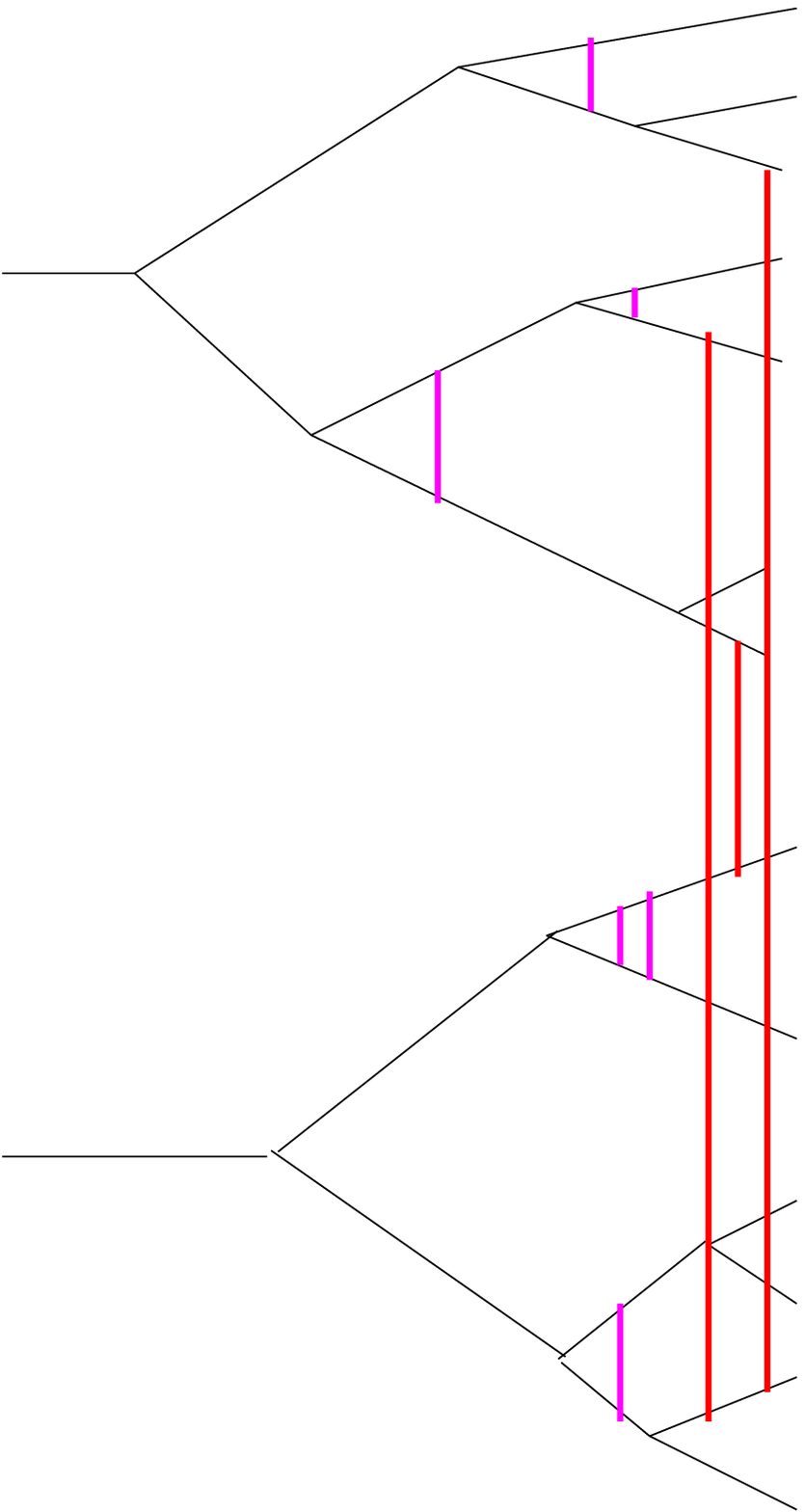,width=0.15\textwidth}
\end{center}
\caption{\label{fig:scat}
Sketch of the scattering of the two evolved hadrons
through splittings only (left)
and through splittings supplemented by nonlinear 
effects (here in the form of internal rescatterings; 
rightmost figure.)
The vertical thick lines would correspond to gluon exchanges
in QCD. Each of them comes with an extra power of $\alpha^2$ which
is not enhanced by corresponding powers of the rapidity.
}
\end{figure}

Our study assumes the following
standard picture of scattering in the QCD parton
model (see Fig.~\ref{fig:scat}): 
An asymptotic hadron, made of valence partons, evolves into
a set of quarks and gluons spread in impact parameter space, when
its rapidity is increased.
The rules of splitting (and recombination when nonlinear effects 
are taken into account in the evolution) 
of the partons are
fixed by the QCD Lagrangian. Two such 
hadrons interact with each other
by exchanging, with probability of the order of
$\alpha^2$ ($\alpha$ is the strong coupling constant),
one or several gluons between the partons of similar
sizes and impact parameters that are 
present in the wavefunctions of the two hadrons
at the time of the interaction.

The outline goes as follows.
In the next section (Sec.~\ref{sec2}), we introduce a model for
partonic state evolution under splittings only.
Sec.~\ref{sec3} is devoted to the formulation of the
unitary scattering amplitudes 
built from these states, that we compare to cases
in which saturation is included explicitly in the evolution.
The numerical study of the many variants of the
toy model, in different frames and within different
unitarization schemes, is the object of Sec.~\ref{sec4}.
In Sec.~\ref{sec5} we present an alternative calculation,
based on the numerical resummation of 
the asymptotic series
in the number of Pomerons that are
exchanged. We state our conclusions and some prospects in 
the last section.


\section{\label{sec2}Toy model for linear parton evolution}

\subsection{\label{sec2A}Construction of the model}

Focussing on one particular impact
parameter, one may reduce the QCD problem to one transverse
dimension only. It is very important to keep the
dynamics in the transverse space if one wants the model
to be representative of some of the physics of QCD.
However, 
we do not aim at incorporating all aspects of QCD: We need
a simple model, that is easy to formulate
in a way which can be implemented in 
the form of a Monte-Carlo event generator.
Let us assume that there be only one type of object in the
theory, which could be gluons or, more accurately, color dipoles,
and that the latter may be fully characterized by a single
``space'' variable, which represents for example their size
in the transverse plane. There is in addition an evolution 
variable: the rapidity. So far, we have in fact 
just idealized the
color dipole model \cite{M1}.

To further simplify the model,
we discretize it both in space and in rapidity.
The evolution rule is the following:
When the rapidity $y$ is increased by one unit,
a particle at position $i$ on the lattice
may be replaced, with probability $\frac12$, 
by two offspring at respective positions
$i+j$ and $i-j$. For the distribution of $j$, 
we choose:
\begin{equation}
\text{Proba}(j)=\left(1-\frac{1}{e}\right)e^{-j}.
\label{probaj}
\end{equation}
This rule obviously leads to an exponential increase
of the number of objects on each site, which can eventually
break the unitarity constraints.
Nevertheless, we do not a priori specify a saturation mechanism
at this level: Several
options of implementing unitarity of the scattering amplitudes
will be examined in the next section, and saturation,
in the sense that the number of particles is prevented from
exhibiting an exponential growth forever,
will only be one of them.

Note that the choice of discrete transverse space and especially
discrete $y$ is not very natural for a model that
is meant to mimic QCD, in particular since discretizing
$y$ obviously breaks boost invariance (which makes sure that the
rapidity evolution can be shared arbitrarily between the
two incoming hadrons without affecting the observables). 
However, this choice is dictated
by technical reasons: We will need to be able to generate
millions of events within some reasonable computer time.

Let us define the generating function for the probability
of the different configurations as
\begin{equation}
Z(y,\{x_k\})=\sum_{\{n_k\}} \left(\prod_k x_k^{n_k}\right) P(y,\{n_k\}),
\end{equation}
where $P(y,\{n_k\})$ is the probability of having
$\cdots n_1,\cdots,n_k,\cdots$ particles
on sites $\cdots 1,\cdots,k,\cdots$
at rapidity $y$.
This function contains all information about the statistics of the
particle numbers on all sites. For example, 
the correlator of the number of particles on sites 1 and 2
is obtained by taking two derivatives:
\begin{equation}
\left.\frac{\partial Z}{\partial x_1 \partial x_2}
\right|_{x_k=1}=\langle n_1 n_2\rangle.
\end{equation}
The generating function $Z$ obeys the evolution equation
\begin{multline}
Z(y+1,\{x_k\})=\frac12 Z(y,\{x_k\})\\
+\frac12\left(1-\frac{1}{e}\right)
\sum_{j=0}^\infty e^{-j} Z(y,\{x_{k-j}\})Z(y,\{x_{k+j}\}).
\label{generating}
\end{multline}
This equation is easily derived by considering the 
very first step in rapidity and with the help of the probability
distribution~(\ref{probaj}).
This is actually a version of what we call in QCD
the Balitsky-Kovchegov (BK) equation \cite{Balitsky:1995ub,K1,K2} 
associated to this simplified model.
The physical $S$-matrix element in this framework, 
$S_\text{BK}(y,i)$, 
is a particular value of $Z$,
obtained from the above equation
by setting $x_{k\ne i}=1$ and $x_i=e^{-\alpha^2}$.
The nonlinearity in Eq.~(\ref{generating}) makes sure that $S$ be
unitary. We will
define the scattering amplitudes properly
only in Sec.~\ref{sec3}, but before, we may
summarize the main known properties of $S_\text{BK}$
seen as a solution of Eq.~(\ref{generating}).

\subsection{\label{sec2B}Basic properties of $S_\text{BK}$}

We know that $S_\text{BK}$ has the form of a wave front that travels
towards larger values of $|i|$ under rapidity evolution.
Its known properties are very well documented, and we refer
the reader to the original papers in QCD \cite{MP1,MP2,MP3} 
(see also Ref.~\cite{MT} for the pioneering calculations,
but without the connection to traveling waves)
or to reviews
in mathematical physics \cite{VS} for the details.
Traveling waves have a phenomenological signature in
high-energy scattering, called ``geometric scaling'' and found
in deep-inelastic scattering data \cite{SGBK}.
We give here the technical features of our particular model
without much justification.

When $i$ and $y$ are large, 
\begin{equation}
S_\text{BK}(y,i)\sim 1-e^{-\gamma_0(i-{\cal I}_y)},
\label{formSBK}
\end{equation}
where ${\cal I}_y$ is the position of the wave front at 
rapidity $y$.
In the context of QCD, ${\cal I}_y$
would be the logarithm of the squared
saturation scale.

As well-known, $\gamma_0$ and ${\cal I}_y$ are determined
from an ana\-lysis of the linearized equation~(\ref{generating}).
We insert
Eq.~(\ref{formSBK}) 
into Eq.~(\ref{generating}), and
after linearization and some easy algebra, we arrive at
the expression for the velocity $V$ of a front
of the form~(\ref{formSBK}) with decay rate $\gamma$:
\begin{equation}
V={\cal I}_{y+1}-{\cal I}_y=\frac{\chi(\gamma)}{\gamma},
\end{equation}
where
\begin{multline}
\chi(\gamma)=
\ln\bigg[
\frac12+\frac12\left(1-\frac{1}{e}\right)\times\\
\times\left(\frac{1}{1-e^{\gamma-1}}+\frac{1}{1-e^{-\gamma-1}}\right)
\bigg]
\end{multline}
is the characteristic function of the evolution kernel
in Eq.~(\ref{generating}).
The relevant value of $\gamma$ at large rapidity is the
one that minimizes $V(\gamma)$.
This minimization gives
\begin{equation}
\gamma_0=0.607187\cdots\ ,\ \
V_0=1.02935\cdots
\end{equation}
which are the decay rate and the velocity
of the front
at infinite rapidity.
The asymptotics is approached
as\footnote{The first term $V_0$ was already discussed
by Gribov, Levin and Ryskin in QCD in the 80's \cite{GLR}.
The second term was around in statistical physics
since some time, see Ref.~\cite{VS}, but was rediscovered
independently by Mueller and Triantafyllopoulos in QCD \cite{MT}.
The third term was computed more recently \cite{VS}, and adapted to
QCD in Ref.~\cite{MP3}.
}
\begin{equation}
V(y)=V_0-\frac{3}{2\gamma_0}\frac{1}{y}
+\frac{3}{2\gamma_0^2}\sqrt{\frac{2\pi}
{\chi^{\prime\prime}(\gamma_0)}}\frac{1}{y^{3/2}}+\cdots
\label{Vy}
\end{equation}
where the dots stand for subleading terms
whose analytical expression is not yet known.

Starting from an initial condition of the form
\begin{equation}
S(y=0,i)=1+(e^{-\alpha^2}-1)\delta_{0,i},
\end{equation}
that is to say, 
\begin{equation}
\begin{split}
&P(y=0,n_{k\ne 0}=0\ \text{and}\ n_{k=0}=1)=1,\\
&P(y=0,\text{all other config.})=0,
\end{split}
\end{equation}
a front is formed (i.e. blackness is reached; 
$S(y,0)\ll 1$)
after a rapidity of the order of
\begin{equation}
y_F=\frac{1}{\chi(0)}\ln \frac{1}{\alpha^2},
\label{yF}
\end{equation}
and it relaxes to its asymptotic shape
up to a resolution\footnote{%
We mean that the front is in its asymptotic 
shape~(\ref{formSBK})
for $S<1-\alpha^2$. Indeed, the asymptotic
shape sets in first in
the vicinity of the black
region, and subsequently diffuses upwards. If one
is not able to resolve details of size greater than $\alpha^2$,
then it is enough that $S$ look asymptotic in the above region.
This distinction is relevant when one goes beyond the BK equation and
one takes into account the fluctuations, and $y_R$ is then
a physical relaxation ``time''.
} $\alpha^2$ after an evolution 
over
\begin{equation}
y_R=\frac{1}{2\gamma_0^2\chi^{\prime\prime}(\gamma_0)}
\ln^2\frac{1}{\alpha^2}
\label{yR}
\end{equation}
units of rapidity.


\section{\label{sec3}
Unitary scattering amplitudes}

\subsection{\label{sec3A}Formulation of scattering}

We consider the Fock state of a particle initially
at position 0 after a rapidity evolution $y$.
Through the evolution, one eventually gets
a system of $\{n_j\}$ particles.
The probability that this system
does not interact with a target consisting in a single particle
at position $i$ simply reads
\begin{equation}
e^{-\alpha^2 n_i}.
\end{equation}
This, of course, is also what we shall call the $S$-matrix
element for forward elastic scattering.
With a target that consists in a set of $\{m_j\}$ particles on the 
sites indexed by $j$, the probability that 
there be no interaction reads
\begin{equation}
\prod_j e^{-\alpha^2 n_j m_j}.
\end{equation}
We have just assumed the complete independence of the
individual scatterings, and that each of them
occurs with probability $\alpha^2$ if the objects in presence
have the same position on the lattice, and with probability
0 if this is not the case.
Note that there is here a difference with the QCD dipole
model since there, the interaction consists in at most
one gluon exchange between each pair of dipoles,
whereas with our choice any number of exchanges
may occur.

Let us consider two particles, initially at respective sites
$0$ and $i$, which evolve
into systems of $n_j(y)$ and $m_{j}(y)$ particles respectively
on site $j$ after a boost at rapidity $y$.
In the Mueller approach \cite{M2,M3}, the $S$-matrix 
for the scattering of these objects
is defined, for low rapidities, as
\begin{equation}
S(y,i)=\left\langle\prod_j e^{-\alpha^2 n_j(\sigma y) m_{j}
((1-\sigma)y)}
\right\rangle
\label{mueller}
\end{equation}
where the average is taken over the realizations of the
two systems. 
The $i$ index is implicit in the r.h.s.: It is related to
the initial condition that leads, after evolution, 
to the system $\{m_j\}$.
$\sigma$ ranges between 0 and 1 and
defines the share of the rapidity between the two systems.
When $\sigma=0$ (or 1: the target and the
projectile may be exchanged, hence there is a symmetry
$\sigma\rightarrow 1-\sigma$ after the average over the realizations), 
the scattering occurs in the lab frame. In this case,
the above formula gives the solution to the Balitsky-Kovchegov equation
if the states evolve linearly: $S=S_\text{BK}$, 
where $S_\text{BK}$ was discussed in Sec.~\ref{sec2B}.
When $\sigma=\frac12$ instead, it 
means that the scattering takes place
in the center-of-mass frame.

By definition, $S$ is unitary, whether or not there
is a limit on the growth of the number of partons that
interact.
So a priori there is no violation of unitarity, and saturation
is not necessary to prevent $S$ from becoming negative.
However, we know that the discussion is more subtle, and
that the $S$-matrix defined like that
violates boost invariance \cite{Mueller:2004sea}.
Nevertheless, for small enough values of $y$, there should be no
need to put saturation effects in the evolution of the states.

Let us discuss quantitatively the expected range in which saturation
effects may be left out.
In these considerations, we follow the discussion
given by Mueller \cite{M2}.
To simplify, let us first
go to the center-of-mass frame, that is, we
set $\sigma=\frac12$.
Imagine that we start from a situation in which
both objects are at rest.
When the two systems are gradually boosted in opposite
directions in order to increase the center-of-mass energy
of their eventual scattering, the probability
of a high number of partons
in each of them increases. At some point, 
when the unitarity limit is reached (as soon as the
total rapidity is larger than $y_F$),
the probability that
there be two or more interactions between them becomes of order 1,
while each of the Fock states 
remains relatively dilute (they contain only
typically $1/\alpha$ particles each), in such a way
that one can still consider their evolution as linear.
Actually, because 
the systems share half of the rapidity,
they become subject to nonlinear effects (in the form
of internal rescatterings, for instance, or recombinations) 
at rapidity $2\times y_F$.
So this sets the limit beyond which multiple interactions
between linearly-evolving Fock states
are no longer sufficient to fully
describe the scattering.
The extension to an arbitrary value of $\sigma$ is straightforward,
and we get the limit
\begin{equation}
y<\frac{y_F}{\max(\sigma,1-\sigma)}.
\end{equation}
The center-of-mass frame ($\sigma=\frac12$) is naturally the
most favorable case, with the highest limit on $y$, while in the
lab frame, the wavefunction evolution 
can be linear only until blackness of the scattering amplitude
is reached.

\subsection{\label{sec3B}Saturation}

So far, we have considered Fock-state evolution 
through splittings only (Sec.~\ref{sec2}).
A unitary $S$-matrix was constructed
from multiple scatterings in the center-of-mass frame
with linearly-evolving
states (Sec.~\ref{sec3A}). 
However, we know that theoretically, this
procedure has a limited validity, that we have estimated
as $y<2y_F$. In order to compare
the results  for scattering obtained in that framework
to what would be obtained in the case of 
saturated states, we need to
introduce a saturation mechanism.

We can imagine different ways of enforcing saturation.
The simplest way consists in forbidding 
new splittings to a site on which
the number of particles has reached the value $1/\alpha^2$.
We will call this method ``veto''.
The resulting model is
close to other models that have been studied before,
e.g. in Ref.~\cite{EGBM}.

However,
in order to get an approximately boost-invariant
$S$-matrix\footnote{Since rapidity is discrete in this model, 
we can only have approximate 
boost invariance.
We will observe the consequences of this incomplete
realization of the symmetry in our numerical results.
}, one needs instead to replace
the splitting probability $\frac12$ at site $i$ by 
$\frac12 e^{-\alpha^2 n_i}$, in such a way
that for small $n_i$ compared to $1/\alpha^2$, 
the splitting probability tends to
the free splitting rate $\frac12$, and for large $n_i$,
splittings are frozen.
This prescription was proposed by Mueller
and Salam \cite{MS}
long ago and revisited more recently \cite{ISAST}.
We will check numerically the approximate boost invariance.
This method will be called ``saturation''.
It is this one that makes sense in a field-theory framework
where of course observables have to be independent
of the frame in which the scattering is observed.
Note that with this choice for the saturation mechanism, 
the resulting model is close to the one
that was extensively studied 
in Ref.~\cite{ISAST}, except for
the discretized rapidity.

To our knowledge, the uniqueness
of the prescription that leads to boost-invariant saturation has not been 
formally proven in models with spatial dimensions such as the one
that we are building. It is also not known whether
the prescription could be slightly modified in a way that approximately
preserves boost invariance, the characteristics
of the amplitude being at the same time
significantly changed.

Finally, we note that the consequences of
boost invariance (or rather of ``projectile-target duality'') 
has been thoroughly
investigated very recently, also at a quite formal and rigorous
level \cite{KL1,KL2}: The transformation
that corresponds to exchanging the scattering particles 
was translated into an operation on the effective
action of the corresponding model.
So we could check the (approximate) boost invariance of
our model with the technology developed in Refs.~\cite{K1,K2}.
(We have not done so: We shall check boost invariance 
later on, but only numerically).

\subsection{\label{sec3C}Expected properties of the $S$-matrix}

If the partonic evolution is linear (i.e. without saturation
effects), and if the scattering takes place in the lab
frame, then $S=S_\text{BK}$, and $S$ exhibits the properties
listed in Sec.~\ref{sec2B}.

We also know that, whatever the saturation mechanism is, 
as soon as the growth of the number of partons on each site 
is limited once it approaches $1/\alpha^2$,
then the traveling wave solution for $S$ is modified 
\cite{BD,MS,BDMM,Panja}.
Essentially, its asymptotic velocity is corrected as follows \cite{BD,BDMM}:
\begin{equation}
V=V_0-\frac{\pi^2\gamma_0\chi^{\prime\prime}(\gamma_0)}
{2 (\ln(1/\alpha^2)+3\ln\ln(1/\alpha^2))^2}+\cdots
\label{velocity}
\end{equation}
This equation contains the first term of an asymptotic expansion
when $\ln(1/\alpha^2)\gg 1$ (It is actually valid
up to ${\cal O}(\ln\ln (1/\alpha^2)/\ln^3(1/\alpha^2)$). 
We do not intend to go to extremely
small values of $\alpha$, thus the exact value of these asymptotics
will not be probed. What is interesting and universal however
is that $V$ is less than $V_0$.

We will be able to
extract more information from our Monte-Carlo, that we can
compare with non-trivial and characteristic
predictions made for reaction-diffusion systems.
In particular, since the whole 
scattering process is stochastic, there
is a dispersion in the position of the front between different
events. It reads \cite{BDMM}
\begin{equation}
\langle {\cal I}_y^2\rangle -\langle {\cal I}_y\rangle^2=
\frac{\pi^4\chi^{\prime\prime}(\gamma_0)}{3\ln^3(1/\alpha^2)}
y+\cdots
\label{variance}
\end{equation}
While again the exact value will not be probed for
it is too much asymptotic, the characteristic feature
of the variance of the position of the front is that it scales
linearly with $y$.
We recall that this linearly growing variance is at the origin of
the phenomenon of ``diffusive scaling'' in the observable amplitude,
which breaks geometric scaling predicted that comes out of 
the solution
to the BK equation.

As for the case when evolution does not include a saturation
mechanism in the Fock states,
it is difficult to figure out a priori what is going to
happen. We will find out in the next section
by performing numerical
simulations of the different variants of the model.


\section{\label{sec4}Numerical study of the toy model}

We have implemented the model 
defined in Secs.~\ref{sec2} and \ref{sec3}
in the form of a Monte-Carlo
event generator.
There was no major difficulty to be overcome:
The techniques that we have used are standard and the code 
can easily
be reproduced by the reader.

The evolution starts with one particle on a given site
in each of the two systems.
We evolve one or the other systems by steps of one unit in $y$.
We measure the position 
of the traveling wave front by searching,
at each rapidity $y$, the
rightmost site $i_F$ for which $S(y,i_F)<0.5$.
We then define the front position ${\cal I}_y$ from a linear
interpolation between $i_F$ and $i_F+1$.

Let us move on to the results.

\subsection{$S$-matrix}

We compute the $S$-matrix
in 3 different variants of the model: bare multiple scatterings
in the lab frame (which is the BK assumption; $\sigma=0$),
the same but in the center-of-mass frame
(which is Mueller's unitarization procedure; $\sigma=\frac12$)
and multiple scatterings off boost-invariant saturated wave functions
in the center-of-mass frame ($S$ should be boost invariant in this
case: We will check it later).

Tuning the frame in the Monte-Carlo is technically
easy: It is enough
to share the rapidity evolution of the projectile and of
the target
proportionally to $\sigma$.

\begin{figure*}
\epsfig{file=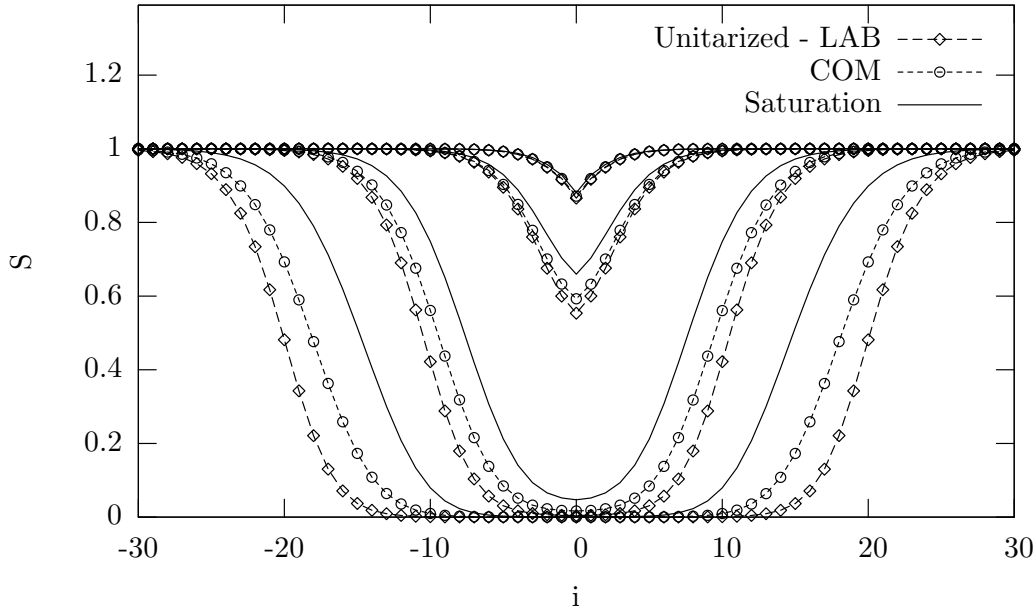,width=0.8\textwidth}
\caption{\label{fig:s_20}
$S$-matrix element as a function of the spatial
coordinate for $y=5,10,20,30$
(from the center of the figure towards the outskirts).
$1/\alpha^2=20$.
For low rapidities ($y=5$ and $y=10$), the evolution is linear (of BFKL
type) and the number of particles grows exponentially.
At $y=20$, the unitarity limit has been reached in all calculations.
Later ($y=30$), the traveling waves are formed and propagate outwards.
Geometric scaling violations are seen in the tilting of the fronts,
except for BK-type scatterings.
}
\end{figure*}

The results are presented in Fig.~\ref{fig:s_20}.
We see that rapidity evolution starts with the blackening of the
central region around $i\sim 0$. At $y\sim 15-20$ two traveling waves form,
and propagate symmetrically
towards larger (resp. smaller) values of $i$.
In the following, our comments will always refer
to the traveling wave that propagates along
the positive $i$ axis.

We see that the fastest wave is observed in the BK model, while
the saturation model gives rise to the slowest one.
We also observe that all waves get slanted during
their propagation, except in the BK model.
This is ``diffusive scaling'', while the BK solutions exhibit
geometric scaling.

In the next paragraphs, we go deeper into the analysis of this
calculation by focussing on the statistics of the position of the front:
Its mean velocity and its variance, for the different variants of the
model.

\subsection{BK equation}

\begin{figure}
\epsfig{file=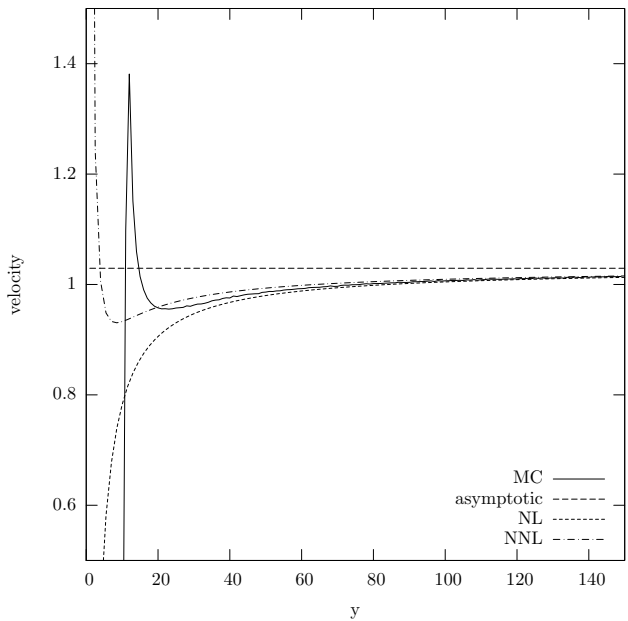,width=0.5\textwidth}
\caption{\label{fig:bk}Front velocity
for the solution of the equivalent BK equation
as a function of
the rapidity, compared to the theoretical
expectations: the different theoretical curves
represent 1,2 or 3 terms of Eq.~(\ref{Vy}).
$1/\alpha^2$ is set to 20.
}
\end{figure}

We go to the lab frame, that is, we put all the
rapidity evolution in one of the objects, while the other
one is at rest.
Multiple scatterings off a system that does not
saturate gives a $S$-matrix that solves the BK equation,\footnote{%
Several groups have solved the exact 
BK equation numerically
in QCD, see for example \cite{AB,GBMS,EGBM}.
A good code ({\tt BKsolver})
is publicly available at {\tt http://www.isv.uu.se/\~{}enberg/BK/}.
}
and hence
that has the form of a traveling wave whose characteristics
are given in Sec.~\ref{sec2B}.
The velocity of the front is 
presented in Fig.~\ref{fig:bk} for $1/\alpha^2=20$.
In order to interpret the results, it is useful
to compute the numerical values of $y_F$ and $y_R$ in 
Eq.~(\ref{yF}) and~(\ref{yR}):
\begin{equation}
y_F\simeq 7.4,\ y_R\simeq 2.2\ \ \text{for}\ 
1/\alpha^2=20.
\end{equation}
We see in Fig.~\ref{fig:bk} that first the velocity
is 0, which corresponds to the phase in which
the parton numbers are building up through a linear
evolution.
(The linear phase is named in QCD after Balitsky, Fadin, Kuraev
and Lipatov (BFKL) \cite{BFKL}).
Later, after about 10 units of rapidity 
(which is the order of magnitude of $y_F$), a sharp peak appears and
decays over a few units of $y$.
This peak may be interpreted as follows: In the initial stage 
of the evolution, when the front is building up, its shape
is steeper than the asymptotic shape~(\ref{formSBK}) in the region
where the position is measured ($S\sim 0.5$), and hence
its velocity is larger than $V_0$. 
Then, the front relaxes to its asymptotic shape in that region,
which takes of the order of $y_R$ steps of rapidity.
In the final stage, for $y\gg 10$, 
the asymptotic velocity is approached in an algebraic way,
in good agreement with the theoretical expectations of 
Eq.~(\ref{Vy}).

\subsection{Front velocity in different models}

\begin{figure}
\epsfig{file=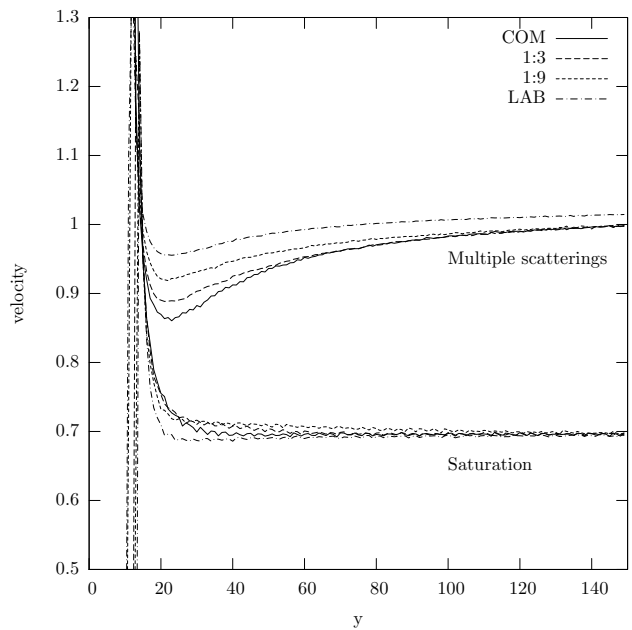,width=0.5\textwidth}
\caption{\label{fig:v20_1}Front velocity
for different unitarization schemes: Multiple scatterings 
(upper bunch of curves), and boost-invariant saturation
(lower curves).
Four distinct frames are considered: $\sigma=0.5, 0.25, 0.1,0$, respectively 
denoted by ``COM'', ``1:3'', ``1:9'' and ``LAB''.
$1/\alpha^2$ is set to 20 in this figure.
}
\end{figure}

We want to compare bare unitarization through multiple scatterings
in different frames to 
the variant of the model in which saturation is
included in a boost-invariant way.
The front velocity for different schemes
and $1/\alpha^2$ set to 20 
is shown in Fig.~\ref{fig:v20_1}.

Bare multiple scatterings lead to a dip around $y\sim 20$
whose depth is maximum for $\sigma=\frac12$.
At large $y$, the velocity $V_0$ is reached algebraically, whatever
the frame is.
At any finite $y$, this model
gives rise to different front solutions in different frames:
Boost invariance is manifestly badly broken.

Solutions with saturation in the evolution
look like expected: The velocity rapidly reaches 
a plateau, with a value that is lower than that of the asymptotic
BK velocity. Theoretically, this should happen after typically $y_F+y_R$
units of rapidity.
For low values of $1/\alpha^2$, the formula giving $y_R$
is expected to need large corrections,
and so the numerical value of $y_R$ is not trustable.
The value of the asymptotic velocity is equal in all frames,
and except for the lab frame, all curves superimpose
reasonably well during the whole rapidity evolution.
We deduce that boost invariance is quite well preserved with our
saturation solution, although not perfectly. We attribute
the lack of complete superposition of the curves to the
explicit breaking of boost-invariance
due to our discretization in rapidity.

\begin{figure}
\epsfig{file=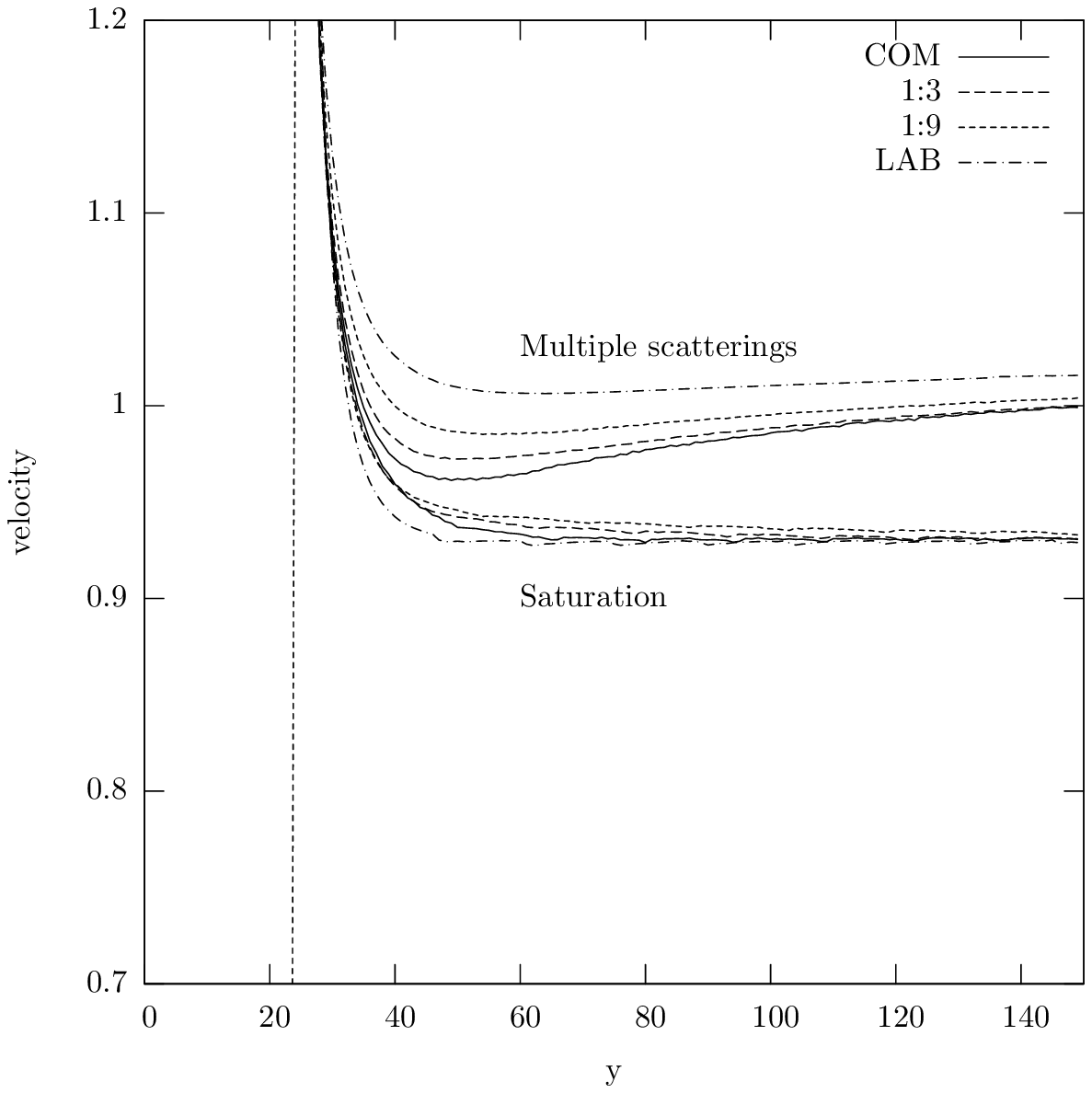,width=0.5\textwidth}
\caption{\label{fig:v2000_1}The same as in Fig.~\ref{fig:v20_1},
but for $1/\alpha^2=2000$.
}
\end{figure}

\begin{figure}
\epsfig{file=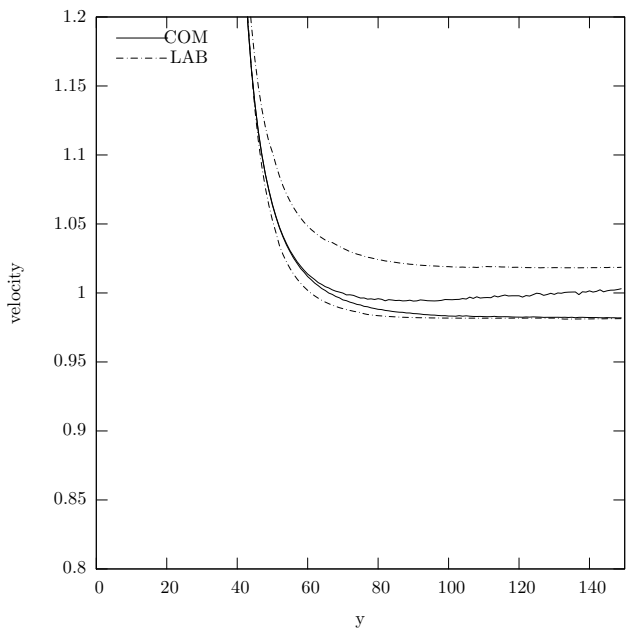,width=0.5\textwidth}
\caption{\label{fig:v200000_1}
The same as in Fig.~\ref{fig:v20_1}, but for
$1/\alpha^2=2\times 10^5$
and only two frames (center-of-mass and lab)
are used.
}
\end{figure}

The same comments are true for larger values of $1/\alpha^2$, see 
Fig.~\ref{fig:v2000_1} and~\ref{fig:v200000_1}.
It is useful to compute $y_F$ and $y_R$ also in that case:
\begin{equation}
\begin{split}
y_F\simeq 18.7,\ y_R\simeq 17.0\ \ &\text{for}\ 1/\alpha^2=2\times 10^3,\\
y_F\simeq 30.1,\ y_R\simeq 43.9 \ \ &\text{for}\ 1/\alpha^2=2\times 10^5.
\end{split}
\label{valuesyFyR}
\end{equation}
We check that a non-zero velocity appears for $y>y_F$.
Actually, $y_F$ given in~(\ref{valuesyFyR}) systematically underestimates
the actual value of the rapidity at which a front is formed.

So far, we see in 
Fig.~\ref{fig:v20_1},\ref{fig:v2000_1},\ref{fig:v200000_1}
that the characteristics of the front agree in the different
models formulated in the center-of-mass frame
only in a very small
interval of rapidity, roughly consistent with the theoretical estimate
$y<2y_F$. Except for large values of $1/\alpha^2$, the front velocity 
in the boost-invariant scheme is not reached by bare multiple scatterings
in the center-of-mass frame.

It is instructive to also study the variance of the position
of the front in the different variants of the model: We analyze this
quantity in the next paragraph.

\subsection{Variance of the front position}

\begin{figure}
\epsfig{file=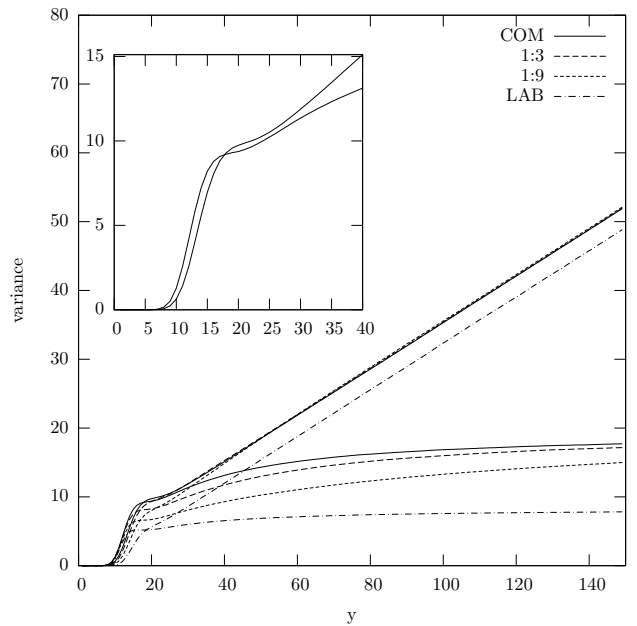,width=0.5\textwidth}
\caption{\label{fig:v20_2}Dispersion of the position
of the front. The curves that converge
to a straight line at large $y$ correspond to
Fock states evolved with a kernel that includes boost-invariant 
saturation, while the ones that flatten correspond
to multiple-scattering unitarization.
$1/\alpha^2$ is set to 20.
{\em Inset:} Zoom into the region of low rapidity for the calculations
in the center-of-mass frame.
}
\end{figure}

\begin{figure}
\epsfig{file=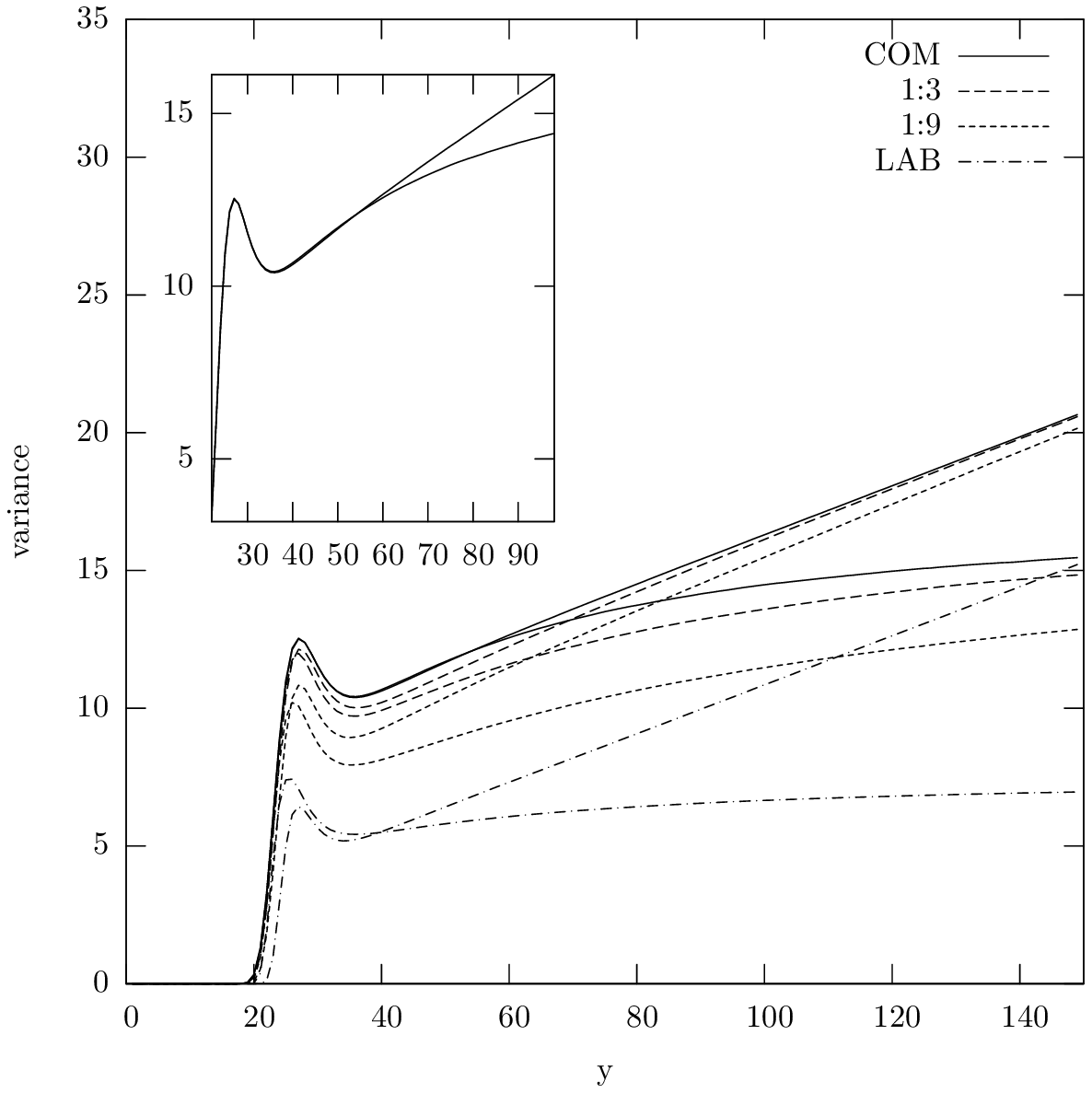,width=0.5\textwidth}
\caption{\label{fig:v2000_2}The same as in Fig.~\ref{fig:v20_2},
but for $1/\alpha^2=2000$.
}
\end{figure}

\begin{figure}
\epsfig{file=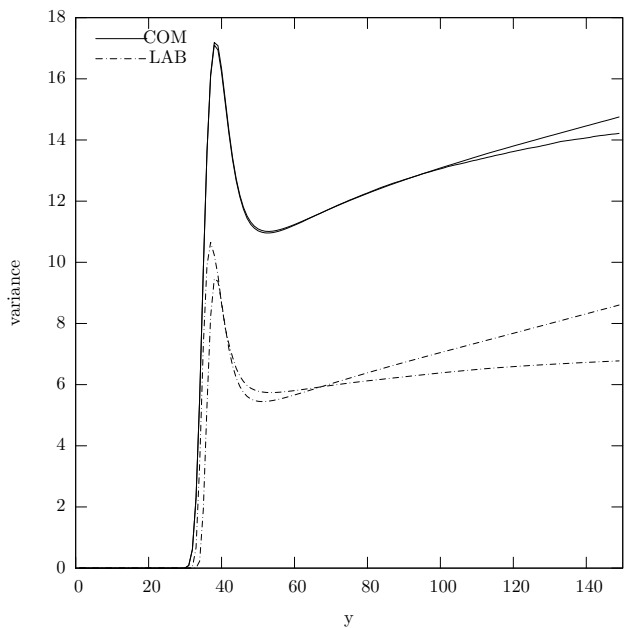,width=0.5\textwidth}
\caption{\label{fig:v200000_2}
The same as in Fig.~\ref{fig:v20_2}, but for
$1/\alpha^2=2\times 10^5$.
}
\end{figure}

We now turn to commenting on the variance of the
position of the front. This is shown in 
Figs.~\ref{fig:v20_2},\ref{fig:v2000_2},\ref{fig:v200000_2}
for different values of $1/\alpha^2$.

First, we notice a sharp difference between saturation
and bare multiple scatterings.
The saturation scheme leads to linearly increasing fluctuations as soon as
the front is formed, in qualitative agreement with Eq.~(\ref{variance}). 
By contrast, the bare multiple-scattering schemes
lead to fluctuations that slow down with $y$.
However, there is a range in rapidity in which
the variances agree in the different models and in all
frames, except for the lab frame which has systematically
less fluctuations. The agreement is particularly
striking at large $1/\alpha^2$.
The range in which the models match 
can be estimated as
$y<2 y_F$ for the center-of-mass frame, 
if $y_F$ is taken to the
rapidity at which the front is effectively formed 
(around the local maximum in the variance) rather than
the numerical estimate done before.

The fact that the calculations in the lab frame
lead to different front velocities and quite different
fluctuations
should maybe not come as a surprise, since
the way in which we treat this frame is very special:
Indeed, we do not allow at all fluctuations in
one of the incoming objects, which is quite unphysical
(quantum objects should be allowed to fluctuate
even if they have a vanishing rapidity),
and which is likely to reduce the event-by-event fluctuations
that are observed in the scattering of the objects.

As was recalled in Sec.~\ref{sec3C},
the linear growth of the variance of the front position
is characteristic of reaction-diffusion
processes, and is very well seen in the model with
saturation,
already for moderately small rapidities. 
The fact that this linear behavior is reproduced
by multiple scatterings in the center-of-mass frame
(see especially Fig.~\ref{fig:v2000_2}) over some
finite range of validity
may be some evidence in favor of the reaction-diffusion
interpretation of high energy scattering.
But admittedly, this linear behavior is seen in a very
small range for small $1/\alpha^2$. On the other hand,
for large $1/\alpha^2$, $y_R>y_F$ and thus the front
has not properly relaxed before genuine saturation effects should
be taken taken into account, for $y\sim 2y_F$.
As a matter of fact, we see that in the range in 
which the models agree in the case $1/\alpha^2=2\times 10^5$,
the asymptotic slope of the variance has not been reached yet.
So except for a model in which $y_R\ll y_F$ and $y_F\gg 1$,
which are two conditions that are difficult to realize
in actual models,
it is very difficult to make convincing statements
on the ability of bare multiple scatterings
to mimic a reaction-diffusion process.

So far, we have considered boost-invariant saturation only.
Boost invariance is a basic requirement for
the model to be consistent with field theory.
We wish however to compare to a non boost-invariant
scheme.

\subsection{Other (non boost-invariant) saturation scheme}

We adopt the alternative scheme of saturation
which consists in vetoing further particle splittings
to a site as soon as the number of particles on this very site
has reached the value $1/\alpha^2$ (see Sec.~\ref{sec3B}).
The small-$\alpha$ asymptotics of the statistics of the front position
should be insensitive to the exact way how saturation occurs.
However, subleading effects at finite $\alpha$
have no reason to be identical.

\begin{figure}
\epsfig{file=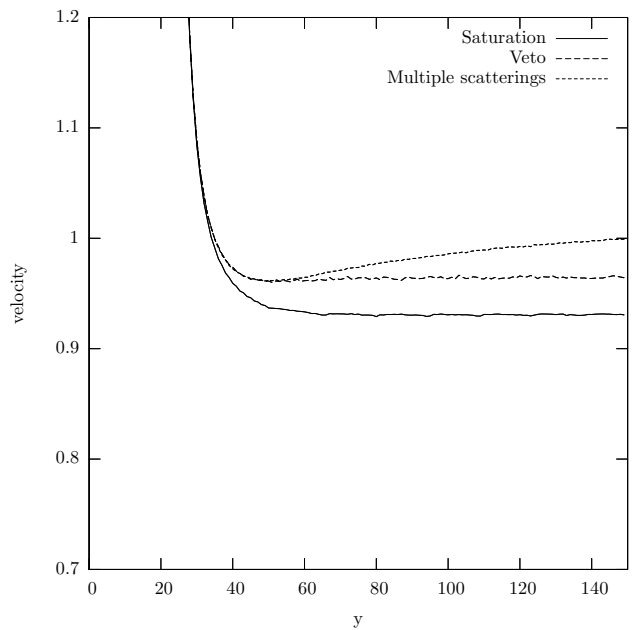,width=0.5\textwidth}
\caption{\label{fig:veto2000_1}Comparison
of the front velocities obtained with
a linear evolution and multiple-scatterings in the COM
system, and with the two different ways
of adding parton saturation 
(veto and boost-invariant saturation).
$1/\alpha^2=2000$.
}
\end{figure}

We see indeed in Fig.~\ref{fig:veto2000_1} that the asymptotic
velocity is higher for this scheme than for boost-invariant
saturation. Moreover, multiple-scattering unitarization
leads to a velocity that, at low $y$, is closer to the one
obtained from the veto procedure.

\begin{figure}
\epsfig{file=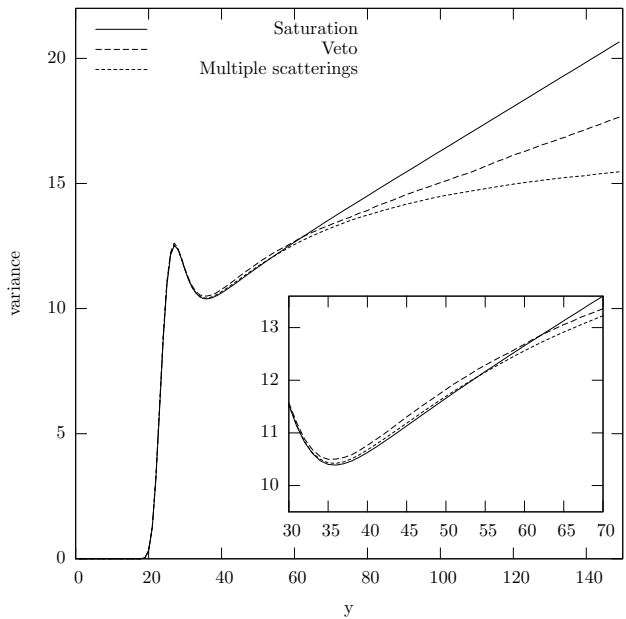,width=0.5\textwidth}
\caption{\label{fig:veto2000_2}
The same as in Fig.~\ref{fig:veto2000_1}, but now the
variance of the position of the front is shown instead
of the velocity. {\em Inset:} zoom into the region in which
the three calculations almost agree.}
\end{figure}

However, if one observes the variance of the position of the front 
(Fig.~\ref{fig:veto2000_2}),
one sees that the large-$y$ slope is quite different between the veto
scheme and the boost-invariant scheme.
Around $y\sim 50$, it is clear 
that the unitarization scheme is
much closer to boost-invariant saturation.
On the other hand,
the large-$y$ slope of the variance in the boost-invariant
scheme looks like the continuation of the slope
in the domain in which they agree.

In this sense, 
boost-invariant saturation
is the natural continuation of the multiple-scattering unitarization
procedure, since it ``knows'' about the fluctuations of saturated
Fock states at large values of the rapidity.
The whole difficulty would be 
to find how, in practice, to perform this continuation.

\section{\label{sec5}
Alternative calculation: Resummation of asymptotic series}

\subsection{Pomeron-loop expansion}

In the framework of the approximations
of high-energy scattering that we are considering
(multiple scatterings of unsaturated Fock states
in the center-of-mass frame), 
the scattering cross section may be computed from
Eq.~(\ref{mueller}).
That expression may also be further expanded in powers of
$\alpha^2$. We get the following series:
\begin{equation}
S(y,i)=\sum_{k=0}^\infty
\frac{(-\alpha^2)^k}{k!}
\left\langle
\left[\sum_j n_j(y/2) m_{j}(y/2)\right]^k
\right\rangle.
\label{Sasymptotic}
\end{equation}
$k$ is the number of Pomerons exchanged between 
the target and the projectile:
$k=1$ is the tree-level (BFKL) term,
$k=2$ is a one-loop contribution, and so on.

The series that has been obtained is actually
a divergent series: The term of order $k$
behaves like $k!$ because essentially, 
$\langle n^k\rangle\sim k!$.
The problem is now the following:
Assuming that we know the first terms in this divergent 
series,
can we get an estimate of the fully resummed series?

\subsection{Overview of the resummation method}

We have to perform an infinite sum, 
where only a fixed number of terms is known. 
There are different possibilities to estimate the limit of 
this sum based on these restricted pieces of information. 
Depending 
on the asymptotic behavior of the partial
sums,
literally summing the first 
elements of a series in many cases is too slow or even does
not converge, as in our case.
In physics, 
the best known 
techniques to deal with divergent series 
are Borel summation \cite{Borel:1899} and 
Pad\'e approximation \cite{Pade:1892}.

On the other hand, there exists nowadays an enormous number
of non-linear sequence transformations which can 
accelerate the convergence of convergent series and 
which have also shown to be very efficient in summing 
divergent series. Particularly suited is a class of such 
transformations introduced by Levin  \cite{Levin:1973}. 
For an introduction to the topic we refer the reader to 
Ref.~\cite{Homeier:2000,Caliceti:2007ra} and references therein.
Here we just sketch the main formulae.

Consider the partial sums $ s_n $ with limit 
(or ``antilimit'', which is how the resummed value of
 a formally divergent sum is called)
$s$ and remainder $R_n$:
\begin{equation}
  s_n = \sum_{i=0}^n a_i = s + R_n .
\label{eq:sum-sum}
\end{equation}
The aim now is to find a new sequence of partial sums 
$s'_n$ such that $R'_n/R_n \to 0$  for $n\to\infty$.
An important building block of a specific sequence 
transformation is  a {\it remainder estimate} $\omega_n$ which 
should reflect the behavior of the exact remainder.
Since the remainder estimate only describes the leading 
behavior of the exact remainder, we write
\begin{equation}
  R_n = \mu_n \omega_n ,
\label{eq:sum-remainder}
\end{equation}
where $\mu_n$ is of the order of 1 and converges to some 
(unknown) number. The second specific ingredient is a set 
of functions $\psi_i(n)$ on which the $\mu_n$ are decomposed. 
For the original Levin-transformation \cite{Levin:1973} the 
set $\psi_i(n)=(n+\beta)^{-i}$ was chosen, where $\beta$ is 
an arbitrary real positive parameter which in general is set to 1
(We will however choose a different value of $\beta$ in our application
of the method). We write
\begin{equation}
  \label{eq:sum-mu}
  \mu_n \sim \sum_i^\infty c_i \psi_i(n) \quad n\to\infty.
\end{equation}
Of course, the coefficients $c_i$ are unknown, but if we 
truncate the sum in Eq.~\eqref{eq:sum-mu}, we can interpret 
Eq.~\eqref{eq:sum-sum} as a model sequence
\begin{equation}
  \label{eq:sum-model}
  \sigma_m = \sigma + \omega_m\sum_{i=0}^{k-1} c_i\psi_i(m) .
\end{equation}
Inserting the values of the partial sums at hand for $\sigma_m$, 
we have for $m\in \{0,1,\ldots,k\}$ a system of linear equations 
which can be solved exactly for $\sigma$ by Cramer's rule. 
By a recursive approach one can deduce a compact expression 
for $\sigma$ which circumvents the evaluation of large determinants:
\begin{equation}
  \label{eq:sum-masterformula}
  \sigma = \frac{\sum_{i=0}^{k-1} \lambda^{(k-1)}_{0,i} 
\frac{s_{i}}{\omega_{i}}}{\sum_{i=0}^{k-1} 
\lambda^{(k-1)}_{0,i} \frac{1}{\omega_{i}}} ,
\end{equation}
where the coefficients $\lambda^{(k)}_{n,i}$ have
to be calculated for a given
set of functions $\psi_i(n)$ and are independent of 
the concrete remainder estimate.
At this general level, rigorous mathematical statements about the
convergence are hardly possible, but for well-posed expansions as they
appear in physical problems $\sigma$ converges to $s$ when $k\to\infty$
\cite{Weniger:1989}.

For the remainder estimate, Levin introduced three variants 
(see Tab.~\ref{tab:sum-remainder}). The $t$-variant is 
adapted for the case of linear convergence, while the 
$u$- and $v$-variants shall be also usable for logarithmic 
convergence. The $d$-variant has been proposed in 
Ref.~\cite{Smith:1979} especially for alternating 
logarithmically convergent series. The $c$-variant has been 
designed for the same purpose \cite{Caliceti:2007ra}. All of 
them have been used for the summation of divergent sums as 
well. In face of these preliminary considerations, there is 
no strict rule which remainder estimate one has to use. For 
our problem, it turned out that variants $c$ and $d$ are 
less successful to the other ones, where $u$ and $v$ give the most 
stable results.

\begin{table}[h]
  \begin{center}
  \begin{tabular}{c||c|c|c|c|c}
    Levin-type    & $t$ & $u$ & $v$ & $d$ & $c$ \\
\hline
    $\omega_n$ & $a_n$ & $(n+\beta)a_n$ & 
$\frac{a_na_{n+1}}{a_n-a_{n+1}}$ & $a_{n+1}$ & 
$\frac{a_na_{n+2}}{a_{n+1}}$
  \end{tabular}
  \end{center}
  \caption{Remainder estimates  for different sequence transformations.}
  \label{tab:sum-remainder}
\end{table}

The convergence significantly improves if one compiles 
inverse factorials (or more accurately Pochhammer symbols) 
instead of inverse powers to the function set $\psi_i(n)$ 
leading to the Weniger $\mathcal{S}$- or 
$\mathcal{M}$-transformations \cite{Weniger:1989}. The 
explicit expressions for the functions $\psi_i(n)$ and the 
coefficients $\lambda^{(k)}_{n,i}$ used in 
Eq.~\eqref{eq:sum-masterformula} are given in 
Tab.~\ref{tab:sum-lambda}. They can be combined with the same 
remainder estimates and are then labeled (analog to 
Levin-transforms, see Tab.~\ref{tab:sum-remainder})
 $\tau$, $y$, $\phi$, $\delta$, $\chi$ for 
the $\mathcal{S}$-transformations, and $T$, $Y$, $\Phi$, 
$\Delta$, $X$ for the $\mathcal{M}$-transformations.
 While the $\mathcal{M}$-transformations are of no use for us,
as long as we stay with $\beta=1$,
the $\mathcal{S}$-transformations provide good results with 
the same ranking concerning the remainder estimates, i.e. 
the most stable results are obtained using the $\phi$-transformation.
But the remainder estimate is not pure trial-and-error. 
From the knowledge about our sum it is clear that the 
remainder estimate $t$ correctly describes the behavior of the sum.

What goes wrong? Usually these sequence transformations are 
used when the available elements are known with high precision. 
By contrast, we calculate our elements by a Monte Carlo
 simulation in which the higher moments are afflicted by  
larger relative statistical error than the smaller ones.
Moreover, these higher moments cause a statistical error by their mere
size compared to the other moments when implemented on a computer with
limited precision.
The sequence transformations discussed so far emphasize these 
larger moments assuming that they are already closer to the 
limit. If we refrain from a too large impact of these 
larger moments, we can shift to a more balanced combination 
of the moments by increasing $\beta$ in the coefficients 
$\lambda^{(k)}_{n,i}$. In the limit $\beta\to\infty$ one 
would obtain a symmetric weighting of large and small 
moments $\lambda^{(k)}_{n,i}=(-1)^i \binom{k}{i}$, known 
as the Drummond-transformation \cite{Drummond:1972,Weniger:1989}.

\begin{table}[h]
  \begin{center}
  \begin{tabular}{c||c|c}
    type & $\psi_i(n)$ & $\lambda^{(k)}_{n,i}$ \\
\hline
Levin & $\frac{1}{(n+\beta)^{i}}$ & 
$(-1)^i \binom{k}{i} \frac{(n+\beta+i)^{k-1}}{(n+\beta+k)^{k-1}}$ \\
Weniger $\mathcal{S}$ & $\frac{1}{(n+\beta)_i}$ & 
$(-1)^i \binom{k}{i} \frac{(n+\beta+i)_{k-1}}{(n+\beta+k)_{k-1}}$ \\
Weniger $\mathcal{M}$ & $\frac{1}{(-n-\beta)_i}$ & 
$(-1)^i \binom{k}{i} \frac{(-n-\beta-i)_{k-1}}{(-n-\beta-k)_{k-1}}$ 
  \end{tabular}
  \end{center}
  \caption{$\lambda$-coefficients for different sequence transformations.}
  \label{tab:sum-lambda}
\end{table}

Finally we use the remainder estimate of the Weniger 
$\tau$-variant with $\beta=100$.

Beside these Levin-type transformation there exist also many 
other schemes which are less generally applicable. They cannot 
be written in the form of Eq.~\eqref{eq:sum-masterformula} 
but are given by a recursive definition. From these we also 
tried the $\epsilon$ algorithm \cite{Wynn:1956} and thereby 
Shanks transformation \cite{Shanks:1955}, the Aitken $\Delta^2$ 
process \cite{Aitken:1926} in its iterated form 
\cite{Wimp:1981,Weniger:1989}, the $_p\bf{J}$ transformation 
\cite{Homeier:1995}, Brezinski's $\theta$ algorithm 
\cite{Brezinski:1971} and its iterated formulation \cite{Weniger:1989}.

\subsection{Numerical implementation and results}

We wish to apply the resummation methods described above to the computation
of $S$ in Eq.~(\ref{Sasymptotic}).

We need to compute numerically the first few terms of the series
(\ref{Sasymptotic}).
These are actually proportional to the 
moments of the one-Pomeron exchange
amplitude, which reads
\begin{equation}
\alpha^2 \sum_j n_j(y/2) m_{j}(y/2),
\label{1pom}
\end{equation}
where $\{n_j\}$ and $\{m_j\}$
are realizations of
systems of particles (linearly) evolved by the Monte Carlo
algorithm described above.

The numerical implementation of this calculation
is straightforward. For each event,
the one-Pomeron amplitude is evaluated for
all values of $i$ and $y$.
Then, its $k$-th power is computed, and the average
over events is performed.
We then apply the resummation methods described above
to get the result for $S$.

While for the calculations of Sec.~\ref{sec4} one million of events
were enough, here, we need hundreds of millions of them. Indeed,
it turns out that statistical errors are large
for higher moments of Eq.~(\ref{1pom}).

In practice, we generate $5\times 10^8$ events.
We consider the resummation of $k=5,6\cdots$ terms, up to 30,
for different values of $\alpha^2$.
We consider that this is the present technical frontier,
since a few months of calculation were already needed to achieve
this number of events.

\begin{figure*}
\epsfig{file=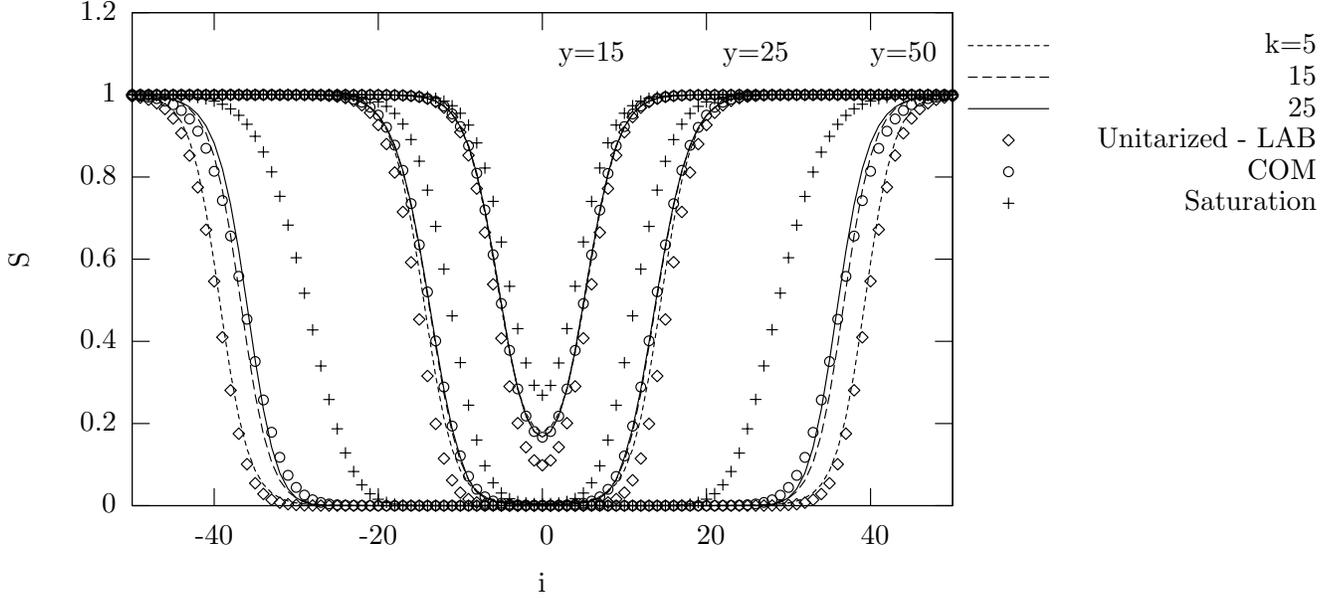,width=1\textwidth}
\caption{\label{fig:S}
Resummed $S$-matrix for 3 rapidities.
Different calculations are compared: multiple scatterings in
the lab and COM frames and boost-invariant saturation
(points). The three curves correspond to the resummation
of the asymptotic series using the Weniger ${\cal S}$-transform
for 5,15 and 25 terms respectively.
}
\end{figure*}

\begin{figure}
\epsfig{file=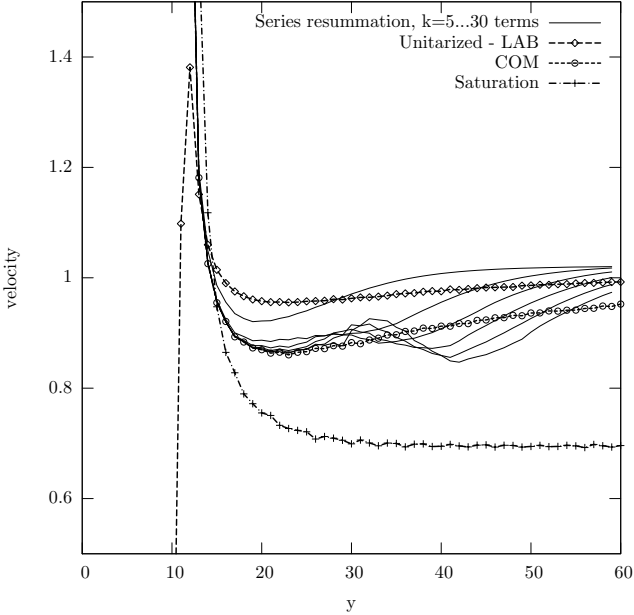,width=0.5\textwidth}
\caption{\label{fig:v20}
Front velocity as a function of the total rapidity
from the resummation of the asymptotic series (bunch of curves in
continuous line: the highest
up corresponds to $k=5$, the lowest to $k=30$)
compared to the different calculations.
$1/\alpha^2=20$.
}
\end{figure}

\begin{figure}
\epsfig{file=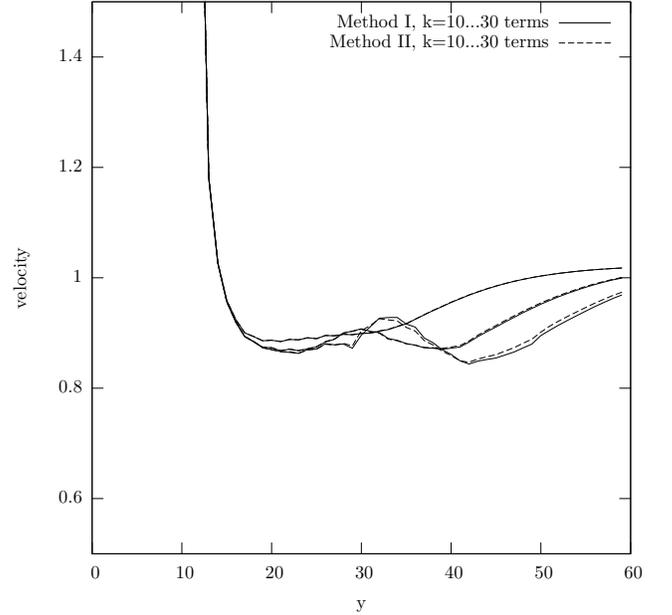,width=0.5\textwidth}
\caption{\label{fig:v20comp}
Comparison of two different resummation methods for the
front velocity: The Weniger ${\cal S}$-transform ($\tau$ version; method I),
and the Weniger ${\cal M}$-transform ($T$ version; method II).
In each case, the parameter $\beta$ is 100.
Low $k$'s lead to higher values of the velocity
in the region $y\sim 20$.
}
\end{figure}

The result of the resummation for $S$ is shown in Fig.~\ref{fig:S}
for 3 different rapidities and for different
number of terms $k$, as a function of the site index.
First of all, we see a good convergence of $S$ when the number of
terms that are taken into account increases.
Second, we see that for $y=15$ and $y=25$, $S$ obtained from
this resummation looks exactly like the one obtained from
multiple scatterings in the center-of-mass frame.
We see that for $y=50$ instead, the result of the resummation does not
coincide with any of the previous calculations.

In order to have a more synthetic estimate of the quality
of the resummation, we can compute the velocity of the front and
compare it to the calculations in Sec.~\ref{sec4}.
We see in Fig.~\ref{fig:v20}
that all calculations match for low $y$.
When more terms are taken into account (larger
value of $k$), the domain in which the statistical and
asymptotic series calculations agree extends towards larger
values of $k$. For $k=30$, the agreement is very good up to values
of $y$ of the order of 25.

\begin{figure}
\epsfig{file=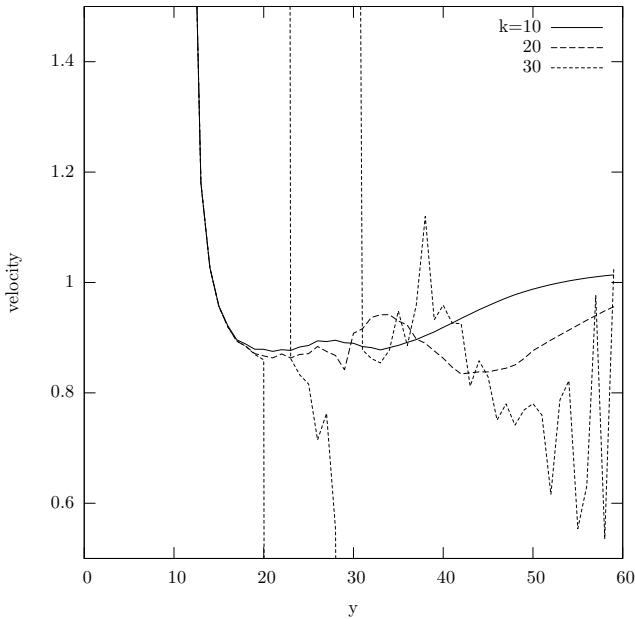,width=0.5\textwidth}
\caption{\label{fig:v20autre}
Resummation using a lower value of the parameter
$\beta$, namely $\beta=10$.
}
\end{figure}

To gain confidence in the stability of our resummation, 
we can compare two out of the many resummation methods described in the
previous section in Fig.~\ref{fig:v20comp}.
We see a very good agreement for all $k$.
We conclude that the discrepancy at large $k$
with the calculation of Sec.~\ref{sec4}
is not due to a failure of the resummation
method, but rather to a lack of the relevant information,
which would be contained in higher-order terms $k>30$.
We would also like to show a method which does not
converge very well: Therefore, we do the same but setting the
parameter $\beta$ to 10 (instead of the higher value 100 that
we have chosen so far). We see in Fig.~\ref{fig:v20autre}
that the resummation for high values of $k$ 
does not give a meaningful result.
More statistics would however help (more events generated, that
is, a better accuracy in all terms).
For yet lower values of $\beta$, such as $\beta=1$, the result would
even be worse.

\begin{figure}
\epsfig{file=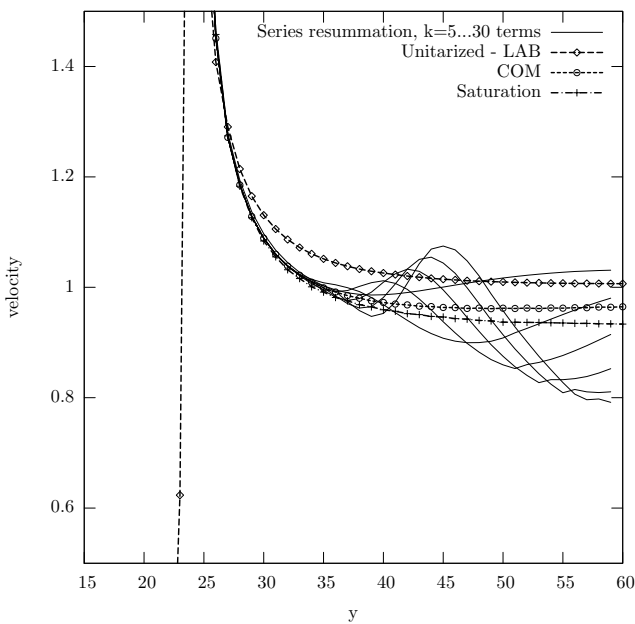,width=0.5\textwidth}
\caption{\label{fig:v2000}
The same as in Fig.~\ref{fig:v20}, but with $1/\alpha^2=2000$.
}
\end{figure}

Finally, we set $1/\alpha^2=2000$, and we see again
the same features, and in particular a good agreement with
the calculations of Sec.~\ref{sec4}, this time up to $y\sim 35$ 
(see Fig.~\ref{fig:v2000}).

Note that we have limited our study to the front velocity
averaged over events, in the statistical language of Sec.~\ref{sec3}.
The variance of the position of the front would require 
a special calculation, that we have not done.


\section{Conclusion and outlook}

The goal of this paper was twofold. 
First, to test the original
unitarization procedure through multiple scatterings, without explicit
saturation,
proposed by Mueller in the context of the color dipole model, in the light
of the new understanding of unitarization gained from the analogy with
reaction-diffusion systems.
Second, to try and reproduce these results from an expansion
in the number of Pomerons that are exchanged.

We have shown empirically that the original formulation of unitarization
by Mueller is successful within the (limited) range of validity that had been
assigned to it.
The traveling waves exhibit
a front velocity that is inferior to the one
that would be expected for a system with no saturation
mechanism at all,
and the event-by-event dispersion in the position of the
front grows linearly with the rapidity, as expected
for reaction-diffusion systems.

We have resummed the multiple scatterings in two ways.
The one that was used for example by Salam in his Monte-Carlo,
which consists in averaging over events the scattering matrix,
and another one which relies on an expansion of the $S$-matrix
in powers of $\alpha^2$.
A priori, it was not completely obvious that these two
ways of computing the unitarized dipole-dipole
scattering cross section
would lead to the same result for the $S$-matrix,
since one sums the defining series in a very different order in
both cases.
But the fact that we eventually get the same answer
is reasonable.

The resummation tools that we have set up 
and tested in the simple toy model
studied here will be useful in the future.
Indeed, if we were able to compute order by order the
unitarity corrections,
either in this toy model or in full QCD \cite{B1992,BE},
we would be confronted to the task of summing
the resulting series, which has the structure of
the one that we have studied.
In general, there is no reason why there should be a simple formulation
like Eq.~(\ref{mueller}) for scattering amplitudes,
and field theory would naturally lead to an asymptotic series
in powers of $\alpha^2$.
We see that for moderately small rapidities and $\alpha$,
resumming the asymptotic series numerically is doable,
although very difficult due to the
large number of events that one has to generate in order
to achieve a sufficient accuracy.

In our next publication, we intend to
systematically study
the corrections to the Mueller formulation of unitarization,
in the framework of our toy model.
We are curious to find out whether boost-invariant saturation
may be obtained through splittings in the Fock states
and scatterings only, that is, without
any explicit reference to a saturation mechanism in the formulation
of the model. This is of course a crucial question for QCD,
since saturation seems so difficult to formulate
in a practical way.

Another prospect would be to find an analytical expression 
for $S$ within the Mueller formulation of unitarization:
Since the result contains information on saturation, 
an analytical expression would help.
But this would 
require the knowledge of all moments of the particle numbers,
which are given by the solution of the equivalent of the
Balitsky hierarchy. Even within simple models, 
such an achievement does
not seem to be at hand yet.

\acknowledgments
We thank Dr. U. Jentschura for helpful discussions about resummation techniques.
Our work is partially supported by
the Agence Nationale de la Recherche (France),
contract No ANR-06-JCJC-0084-02.


\end{document}